\documentclass[fleqn,usenatbib]{mnras}

\usepackage{graphicx}
\usepackage{graphics}
\usepackage{amsmath, amssymb}
\usepackage{multirow}
\usepackage{color}
\usepackage{xcolor}
\usepackage{bm}		
\usepackage[normalem]{ulem}
\usepackage{comment}

\usepackage{float}

\def\deg{{^\circ}}

\def\mathnew{\mathsurround=0pt}   
\def\simov#1#2{\lower .5pt\vbox{\baselineskip0pt  
    \lineskip-.5pt\ialign{$\mathnew#1\hfil##\hfil$\crcr#2\crcr\sim\crcr}}}

\def\'#1{\ifx#1i{\accent"13\i}\else{\accent"13#1}\fi}

\def\aap{{A\&A}}
\def\aaps{{A\&AS}}
\def\araa{{ARA\&A}}
\def\aj{{AJ}}
\def\apj{{ApJ}}
\def\apjl{{ApJL}}
\def\nat{{Nature}}
\def\mnras{{MNRAS}}
\def\apjs{{ApJS}}

\def\pasp{{PASP}}

\title[Footprint of MW spiral arms in \textit{Gaia} EDR3]{Kinematic footprint of the Milky Way spiral arms in \textit{Gaia} EDR3}

\author[L. Martinez-Medina]{Luis Martinez-Medina$^{1}$\thanks{Contact e-mail:\href{mailto:lamartinez@astro.unam.mx}{lamartinez@astro.unam.mx}}\href{https://orcid.org/0000-0002-5749-8255}{\includegraphics[scale=0.5]{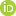}}, Angeles Pérez-Villegas$^{2}$\href{https://orcid.org/
0000-0002-5974-3998}{\includegraphics[scale=0.5]{orcid.png}}, Antonio Peimbert$^{1}$\\
$^{1}$Instituto de Astronom\'ia, Universidad Nacional Aut\'onoma de M\'exico, A. P. 70-264, Ciudad Universitaria CDMX 04510, México\\
$^{2}$Instituto de Astronomía, Universidad Nacional Autónoma de México, Apdo. Postal 106, 22800 Ensenada, B.C., México
}
\date{Released \today}

\pubyear{2021}
\begin{document}
\label{firstpage}
\pagerange{\pageref{firstpage}--\pageref{lastpage}}
\maketitle

\begin{abstract}

The Milky Way spiral arms are well established from star counts as well as from the locus of molecular clouds and other young objects, however, they have only recently started to be observed from a kinematics point of view. Using the kinematics of thin disc stars in Gaia EDR3 around the extended solar neighbourhood, we create x-y projections coloured by the radial, residual rotational, and vertical Galactocentric velocities ($U,\Delta V,W$). The maps are rich in substructures and reveal the perturbed state of the Galactic disc. We find that local differences between rotational velocity and the azimuthally averaged velocity, $\Delta V$, display at least five large-scale kinematic spirals; two of them closely follow the locus of the Sagittarius-Carina and Perseus spiral arms, with pitch angles of 9.12$^{\circ}$ and 7.76$^{\circ}$, and vertical thickness of $\sim400$ pc and $\sim600$ pc, respectively. Another kinematic spiral is located behind the Perseus arm and appears as a distortion in rotation velocities left by this massive arm but with no known counterpart in gas/stars overdensity. A weaker signal close to the Sun's position is present in our three velocity maps, and appears to be associated with the Local arm. Our analysis of the stellar velocities in the Galactic disc shows kinematic differences between arms and inter-arms, that are in favour of Milky Way spiral arms that do not corotate with the disc. Moreover, we show that the kinematic spirals are clumpy and flocculent, revealing the underlying nature of the Milky Way spiral arms.

\end{abstract}                

 
\section{Introduction} 
\label{sec:intro}

Spiral arms are among the most striking structures found in galaxies. More than 60\% of all massive galaxies are spirals \citep{Buta1989,Willett2013} and, with the help of studies of young stellar population and gas, it was proved, in the 1950s, that the Milky Way (MW) is not an exception \citep[e.g.][]{Oort1952,Morgan1953,vandeHulst1954}. In its current picture, our Galaxy contains four major spiral arms \citep[e.g.][]{Georgelin1976,Drimmel2001,Vallee2014}, as well as a strong bar \citep[e.g.][]{Binney1997, Wegg15,Bland-Hawthorn2016}.

According to \citet{Elmegreen1984} and \citet{Elmegreen1990}, spiral galaxies can be classified into three types depending the nature of their spiral arms: i) grand design, with a clearly defined structure and underlying stellar waves, ii) flocculent, characterised by short and patchy spiral arm fragments, and iii) multi-arm, an intermediate type, with a bisymmetric spiral pattern at the centre, and several irregular arms in the outer parts.

Different types of spiral galaxies imply  different formation mechanisms of their spiral arms. While there is no consensus yet on the morphological nature of the Galactic spiral arms, the traditional picture of four major arms as well as some localised features (e.g. the Local arm) suggests that the the MW spiral arms are quite complex with many authors favouring the scenario of flocculent transient spiral structures \citep[e.g.][]{Quillen2002,Quillen2018,Castro-Ginard2021}.

Due to the position of the Sun near the Galactic plane, it has not been possible to obtain the full picture of the structures in the MW disc: interstellar dust and projection effects, as well as the multifaceted characteristics (the makeup) of the arms themselves, interfere with the determination of the spiral arms properties.

During the last two decades, many efforts have being dedicated to increase our understanding of the structure, morphology, and dynamics of the MW spiral arms; going from their location in the Galaxy \citep{Reid2014,Hou2014} to their dynamical influence on the Galactic disc \citep{Hayden2015,Monari2016b,Antoja2018,Martinez-Medina2019}. However, there is still debate on the properties of the spiral arms, e.g., their pattern speed, location, and even their nature \citep{Grand2012,DOnghia2013,Roca-Fabrega2013,DelaVega2015}.

Regarding the locations of the Galactic spiral arms, they have been obtained using a wide variety of techniques and tracers. For example: \citet{Morgan1952,Morgan1953}, using spectroscopic parallaxes of OB stars, detected parts of a spiral arm located between the Sun and the Galactic centre; young open clusters also reside inside spiral arms, and early works measuring their galactic distribution \citep{Becker1963,Becker1970,Fenkart1979} confirmed their role as good tracers. \citet{Dias2005} found how young open clusters must be to trace spiral arms; finding that clusters with ages under $\sim$ 12 Myr are still located inside the spiral arms. More recently, \citet{Castro-Ginard2021} detected overdensities of open clusters younger than 30 Myr and traced the spiral arms with an increased range in Galactic azimuth. 

\citet{Georgelin1976} were the first to propose the existence of four major spiral arms in the MW based on their determinations of the positions of \ion{H}{II} regions and exciting stars. The four-armed nature of the spiral arms has been supported since early numerical works: \citet{Englmaier1999} fitted numerical simulations to the projection of gas dynamics on the CO $l-v$ map, where the best model consists of four major spirals. Recently, these four spiral arms (as well as the Local arm) have been traced and extended using trigonometric parallaxes and proper motions of molecular masers associated to young high-mass stars \citep{Reid2014,Reid2019}. 

Taking advantage of the amount and quality of data available from {\it Gaia}, now it is possible to detect fainter spiral structures than ever before. \citet{Pantaleoni2021} used {\it Gaia} and Alma data for OB stars and found a new structure, which extends diagonally from the Local arm towards the Perseus arm, that they dubbed the Cepheus spur.

Although the Galactic spiral structure has been characterised through the overdensities and increased signals of its different tracers, they are also observable in the kinematics of the stellar and gaseous disc components. It is well known that, as they sweep the disc, spiral arms (along with the bar) trigger anisotropies in the velocities of stars, as can be observed in the solar neighbourhood \citep[e.g][]{Antoja2011,Siebert2012,Hunt2018,Barros2020}. Mapping the kinematics of stars in barred spiral galaxy models, \citet{Roca-Fabrega2014} found that the velocity ellipsoid vertex deviation follows the density peak of the spirals.
Using high quality disc kinematics and distances of red clump stars, \citep{Eilers2020} found the response and location of the Galactic spiral arms.

In this work, we used the full phase-space information based on radial velocities and astrometry (parallaxes and proper motions) available in the Gaia Early Data Release 3 \citep[EDR3;][]{Gaia2021}. We selected a sample of 651,164 disc stars to compute the local deviations from axisymmetric rotation, $\Delta V(R,\phi)=V(R,\phi)- V_{\it avg}(R)$, where $V$ is the rotation velocity and $V_{\it avg}$ is the axisymmetrically averaged rotation velocity. We find that a $x-y$ map coloured by the local $\Delta V$ values displays several large-scale spiral structures. Our analysis shows that some of these kinematic spirals correlate with two of the four MW’s major spiral arms, while some others are clear inter-arm regions.

This paper is organised as follows. In Sec. \ref{sec:data} we describe the data sample. In Sec. \ref{sec:results}, we analyse kinematic maps of the rotational, radial, and vertical components, where we find several large-scale kinematic spirals, including counterparts to the Perseus and Sagittarius-Carina arms. Finally, a discussion and conclusions of our results are presented in Sec. \ref{discussion}.

\section{Data Set} 
\label{sec:data}

We use {\it Gaia} EDR3 that provides 6D phase-space coordinates of 7,209,831 stars, i.e. positions in the sky $(\alpha,\delta)$, parallaxes $\varpi$, proper motions $(\mu_{\alpha}^*,\mu_{\delta})$ \citep{Gaia2021}, and radial line-of-sight velocities \citep{Katz2019} that come from Gaia DR2 \citep{Gaia2018} for stars with $G<$13\,mag. EDR3 represents a significant improvement in both precision and accuracy of the astrometry. \citep{Gaia2021}.

For our sample we restrict ourselves to stars with positive parallax and errors smaller than $20\%$. The distances have been calculated through the inverse of the parallaxes after the zero-point correction \citep[0.017 mas;][]{Lindegren2021}.

In order to convert the positions, parallaxes, proper motions on the sky, and radial velocities of the stars into Galactic phase-space positions and velocities, we use the \texttt{galpy} python tools \citep{Bovy2015}, assuming that: the Sun is located a $R_\odot=8.2$ kpc \citep{Bland-Hawthorn2016,Gravity2019}, a height above the Galactic plane $Z_\odot=0.021$ kpc \citep{Bennett2019}, the velocity of the local standard of rest (LSR) is 236 km s$^{-1}$ \citep{Reid2019}, and the peculiar motion of the Sun with respect to the LSR of $(U_\odot, V_\odot, W_\odot)$ = (11.1, 12.24, 7.25) km s$^{-1}$ \citep{Schonrich2010}. Note: we used Cartesian coordinates for the stellar positions, and polar coordinates for the stellar velocities.

This sample also contains a dynamically hot stellar component that belongs to the Galactic thick disc, unable to trace the mid-plane perturbations. To avoid this contamination we decided to put stringent constraints in the vertical position and velocities, trying to isolate the cold disc component.

We explored different filter sets (starting by not putting any filters at all), in all cases, when studying the kinematic footprint in our sample (see section \ref{sec:DV}), we found signs of the same structures; the specific filter selection we use is the one which presents the clearest signal. That amounts to: $|z|\leq0.25$ kpc, $|U|\leq15$ kms$^{-1}$, and $|W|\leq15$ kms$^{-1}$. Finally, for the rotational velocity, $V$, it is also necessary to avoid outliers, but those stars cannot be identified automatically, as we expect to find structure in the rotation curve (RC); for this we implemented a $1-\sigma$ clipping around the RC as the $V$ filter.

Regarding the remaining 2 spacial coordinates, we restricted ourselves to stars with $-13\leq x\leq-5$ kpc and $-4\leq y\leq4$ kpc, assuming that the Sun's position is (x,y)=(-8.2, 0) kpc. By choosing this set of filters we end up with a subsample that contains 651,164 stars and belongs mainly to the thin disc.

\subsection{Rotation curve fit}
\label{sec:RC fit}

\begin{figure}
\begin{center}
\includegraphics[width=\columnwidth]{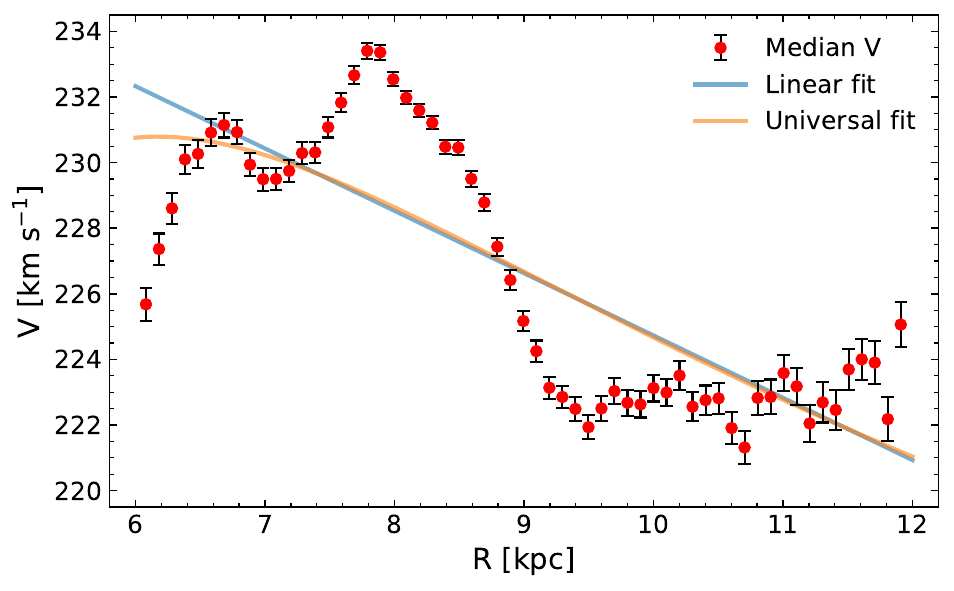}
\end{center}
\caption{Median rotation velocity of our sample as function of the Galactocentric radius (red dots) with error bars. The blue and orange lines represent the best linear fit and the best fit to the universal model, respectively.}
\label{fig:fits}
\end{figure}

Our goal is to characterise the non-circular motions in the stellar kinematics of the extended solar neighbourhood. We start by computing the RC of the sample described in the last section. To this purpose we divide the star sample in 0.1 kpc radial bins and compute the median of the rotation velocities from all stars in each bin. As shown in Fig. \ref{fig:fits} (red dots), the result is a collection of values for the median rotation velocity as a function of the Galactocentric radius. 

The resulting points do not result in a smooth curve. The observed oscillations should not automatically be taken as true features of the RC, since Fig. \ref{fig:fits} represents the sum of the Galactic axisymmetric rotation with the non-circular motions produced by the perturbations in the kinematics of stars. 

These type of features in observed RCs are often associated with an underlying perturbation of the disc, either Galactic or extragalactic \citep{Quillen2009,McGaugh2019,Martinez-Medina2019}, that in turn modifies the kinematics of gas and stars \citep{Shetty2007,Spekkens2007,Baba2016}.

Of particular interest of the observed RC from Fig. \ref{fig:fits} are: the two maxima, at $\sim6.7$ and $\sim7.8$ kpc, and the three minima, at $\sim7$, $\sim9.5$, and $\sim10.7$ kpc. These features are clear non-circular motions due to non-axisymmetric perturbations, and our goal is to quantify how large they are and how they look like in a 2D velocity map, where the non-axisymmetries in velocity will be displayed as a function of Galactocentric radius and azimuth.

To obtain a better contrast of the perturbations in the kinematics we need to subtract the axisymmetric background from the stellar rotation velocities; for this we first obtain an axisymmetrically averaged rotational velocity curve using the data from Fig. \ref{fig:fits} (red dots).

We have assumed that the true RC is smooth, and that most of the contribution to the observed oscillations comes from non-circular motions; if that hypothesis is correct, we expect the locii of the maxima and minima to have a significant dependence on the angle (as will be shown in the following sections). 
  
We fit two simple analytical models: a linear model described by $V(R)= V_0 + \frac{dV}{dR} (R - R_0)$, and a the universal RC as introduced by \citep{Persic1996}, which represents the RC produced by a stellar disc and a spherical dark matter halo. The universal RC is given by:
\begin{eqnarray}
\label{eq:Uni}
    V(x) &=& \sqrt{V_d^2(x) + V_h^2(x)} \\
    V_d^2(x)&=& a_1^2 b \frac{1.97 x^{1.22}}{(x^2 + 0.78^2)^{1.43}} \\
    V_h^2 (x) &=& a_1^2 (1-b)x^2 \frac{1+a^2_3}{x^2+a_3^2},
\end{eqnarray}
where $x= R/R_{\rm opt} = (R/R_0)/a_2$, $a_1$ is the rotation speed at the galaxy optical radius $R_{\rm opt}$, $a_2= R_{\rm opt}/R$, and $a_3$ is the velocity core radius. We adopted $b=0.72$ \citep{Persic1996}. 

In order to completely explore the parameter space, we applied the Bayes' inference through Markov-chain Monte Carlo (MCMC) using the \texttt{python} package \texttt{emcee} \citep[][]{Foreman-Mackey2013}; with this method we obtain the posterior probability distributions. For the linear model fit, we used a uniform prior to restrict the signal of $V_0$ and $\frac{dV}{dR}$. For the universal curve model, we employed Gaussian prior distributions for the $a_1$, $a_2$, and $a_3$ parameters. To represent the final distributions, we adopted the median as the most probable value and the $16^{\rm th}$ and $84^{\rm th}$ percentiles as 1-$\sigma$ uncertainties. The best fit to the linear model has $V_0= 228.15^{+1.49}_{-1.52}$ km s$^{-1}$ and  $\frac{dV}{dR} = -1.90 ^{+0.75}_{-0.77}$ km s$^{-1}$ kpc$^{-1}$ (blue line in Fig. \ref{fig:fits}); while the best fit to the universal model has $a_1 = 230.68^{+2.61}_{-2.12}$ km s$^{-1}$, $a_2=0.82^{+0.10}_{-0.11}$, and $a_3 = 1.30 ^{+0.10}_{-0.11}$ (orange line in Fig. \ref{fig:fits}).

Starting at $\sim7.2$ kpc the two fitting curves are very similar and they are good in describing the overall trend in the outer RC, suggesting either curve could be a good approximation of the ``true" global RC. This comparison shows that, although the universal RC model by \citet{Persic1996} is realistic enough, it is not much more complicated than a simple linear fit. Overall, we prefer the universal curve because it seems to remain a good approximation all the way up to $\sim6.5$ kpc. For our analysis, we adopt the universal curve to represent the underlying averaged azimuthal velocity in our data.

We would like to point out that all theoretical Milky Way RCs have a smooth fall in this disc region, starting at about 6 kpc (i.e. a few kpc beyond the limit of the bar) up to about 15 or 20 kpc (although we are focusing only up to 12 kpc because of the data we have).

\section{Mapping the spiral arms in the stellar kinematics} 
\label{sec:results}

When using the term ``velocity structure" it is sometimes hard to remember that this is not a one dimensional structure; even when we restrict ourselves to the plane, the velocity structure remains a 2-dimensional affair. While using theoretical models, it is possible to construct a galaxy where we can perfectly separate the axisymmetric background from the non-axisymmetric components. Observationally, it is not possible to do that; unfortunately, the rotational velocity signature created by the spiral pattern (or any other non-axisymmetric perturbation) is locally blended with the RC of the axisymmetric components. For the MW, it would only be possible to properly determine the axisymmetric background (and thus isolate the non-axisymmetric component), if we had quality data from the entire Galaxy, or at least of a slab comprising 180 degrees of the Galactic disc; but, when the quality of our data restricts us to distances smaller than about 3 kpc, we only have at our disposal information of $\sim 45^{\circ}$ of the Galactic disc, which makes it easy to confuse radial axisymmetric structures on the RC with the signal produced by the arms (or other non-axisymmetric component). 

If we knew the true global RC, the difference between the local velocities and the RC will be the signal of non-axisymmetric structures. Currently, we do not possess a RC with that quality, but any good approximation (such as the universal RC from sec. \ref{sec:RC fit}) would help us to see those structures.

To further explore the features in the observed rotational velocity, now we account for the azimuthal angle. For this purpose, we denote the radial, rotational, and vertical components of the velocity as $(U,V,W)$, that are positive in the direction of the Galactic centre, Galactic rotation, and north Galactic pole, respectively. Then, we project our stellar sample on the x-y plane and draw a mesh to divide this plane into 0.2 kpc-size cells. 

\subsection{Rotational velocities in the EDR3}
\label{sec:DV}

Although we have three-dimensional information for each star, first we focus on $V$. Furthermore, since our goal is to study the non-axisymmetric features of the rotational velocity, we will use the residual rotational velocity to highlight their signal. For this purpose, in each cell, we will determine the value of the residual rotational velocity
\begin{equation}
\label{eq:DV}
\Delta V(R,\phi)=V(R,\phi)- V_{\it avg}(R) ,
\end{equation}
where $V$ is the median of the rotation velocity of the stars inside the cell, and $V_{\it avg}$ the axisymmetrically averaged rotation velocity, obtained from the universal curve (equation \ref{eq:Uni}), evaluated at the radial position of the cell. This means that at each cell we show a measure of how the rotation in the Galactic plane deviates from the averaged azimuthal velocity.

\begin{figure*}
\begin{center}
\includegraphics[width=\textwidth]{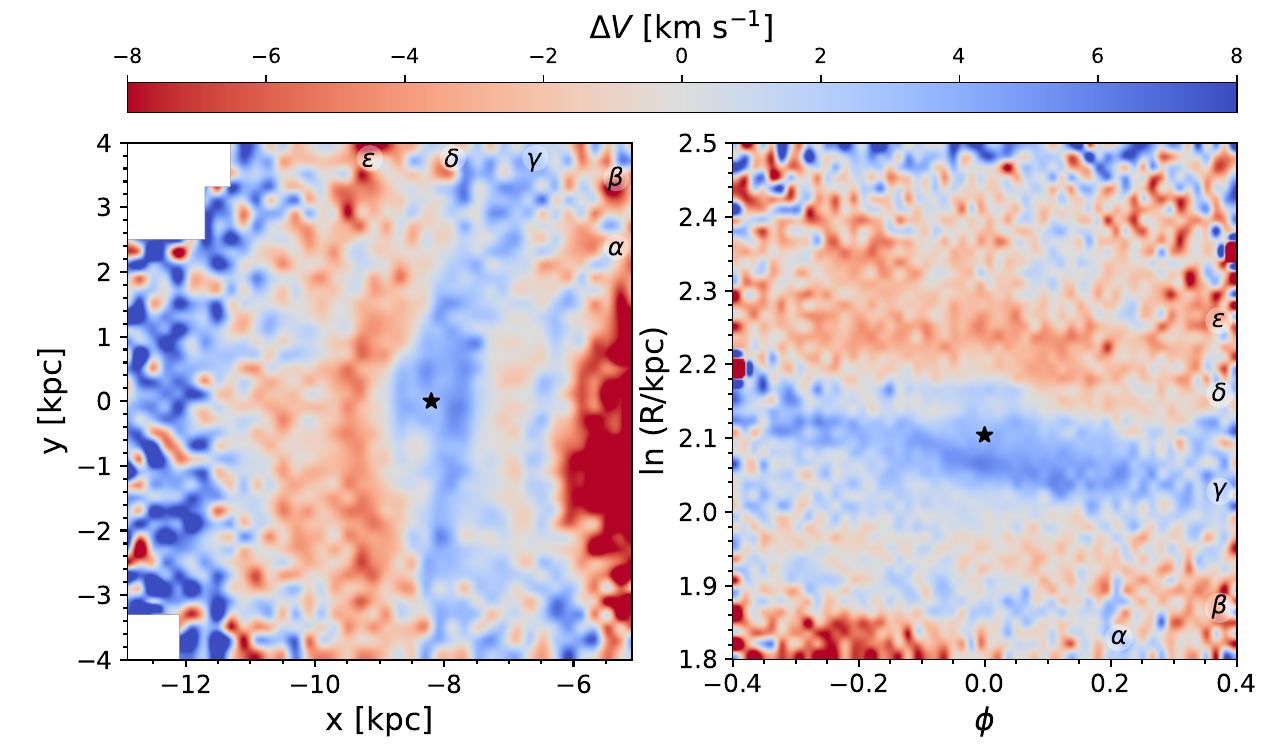}
\end{center}
\caption{Spiral footprint in stellar $\Delta V$ values projected on the Galactic plane. Five spiral structures are labelled from $\alpha$ to $\epsilon$.}
\label{fig:DV}
\end{figure*}

The left panel of Figure \ref{fig:DV} shows the star sample in the x-y plane and the cells coloured with the local value of $\Delta V$. Notice that positive values of $\Delta V$ correspond to regions of the disc where the rotation is faster than the axisymmetric background, while negative values of $\Delta V$ correspond to regions where the rotation is slower. The first result we find is that the kinematics of the disc is full of substructure; in particular several spiral structures are evident.

Since in theory and observations spiral arms tend to behave like logarithmic spirals, it is common to transform the x-y plane to a angle-log plane. On the right panel of Fig. \ref{fig:DV} we plot the same star sample but now in the $\phi-ln(R)$ plane. We notice that here the kinematic spirals appear roughly as straight lines (in this plane a logarithmic spiral will appear as straight line with slope equal to the tangent of the pitch angle of the spiral). 

We defined two structures ($\alpha$ and $\gamma$) by selecting the ``fast" bins (those with $\Delta V>0$) and two structures ($\beta$ and $\delta+\epsilon$) by selecting ``slow" bins (those with $\Delta V<0$). However, the perpendicular profile of the outermost structure shows that it is a double-peaked feature, which we decided to split into two structures, $\delta$ and $\epsilon$. There is an additional red feature inside $R \approx 6$ kpc; this saturated region is likely because the relatively simple structure of the universal RC, which is unable to represent the rich structure of the inner MW disc (near the end of the bar); whatever the specific reason, such discrepancy is beyond the scope of this paper. There is also an outer blue region; but, as it is noisy and boundless, we are not able to characterise the structure in that area.

Because traditional models expect the slow material to accumulate in the arms and the fast material to define the inter-arms  \citep[e.g.][]{Lin1964,Kalnajs1973,Elmegreen1986}; the three red extended structures ($\beta$, $\delta$, $\epsilon$) behave like spiral arms, where the rotation is slower than the axisymmetric velocity; while in blue, with fast rotation, we find two inter-arm-like structures, one marked as $\alpha$, and another that crosses through the Sun’s position, marked as $\gamma$.

The red spirals  display detailed features that show a wide variety of behaviours: while spiral $\delta$ is a single large fragment, potentially part of a larger structure, spiral $\epsilon$ is composed by two main fragments separated by a prominent gap around (-10.2,1.0) kpc, and spiral $\beta$ is fragmented and clumpy, with several small gaps along its extension. Overall, all three spirals show evidence of flocculence; it should be noted that this flocculence is not the more usual density flocculence, but rather is a ``kinematic flocculence", a counterpart that is necessary for the density flocculence to appear.

These results are mostly independent of the selected $V_{\it avg}(R)$; any analysis using a smooth axisymmetric slowly decreasing curve will find the same structures that we are finding: any moderate height or tilt variation (i.e. a reasonable variation) will, at most, produce noticeable changes in the width of the kinematic structures, but would produce only small modifications of the locus of the maxima and minima, and nearly no modifications in the measured pitch angles. Even major changes in the absolute value or inclination of the reference curve could eliminate one (or more) of the structures we have defined (i.e. red to blue or vice versa) but their signal will still be present in the form of a local minimum or maximum, including the same pitch angle that we are observing.

\subsection{Characterisation of the kinematic spiral structures} \label{sec:charact}

To estimate the parameters that define the kinematic spirals detected in the $x-y$ projection of Fig. \ref{fig:DV}, we fitted analytical spirals to these structures, adopting a log-periodic spiral defined by
\begin{equation}\label{eq:spiral}
    \ln (R/R_{ \rm ref}) = - \phi \, \tan \psi,  
\end{equation} 
where $R$ is the Galactocentric radius at a Galactocentric azimuth $\phi$ (defined as 0 towards the Sun and increasing in the direction of the Galactic rotation) for a spiral with a radius $R_{\rm ref}$ at reference azimuth zero, and pitch angle $\psi$. 

\begin{figure*}
\begin{center}
\includegraphics[width=\textwidth]{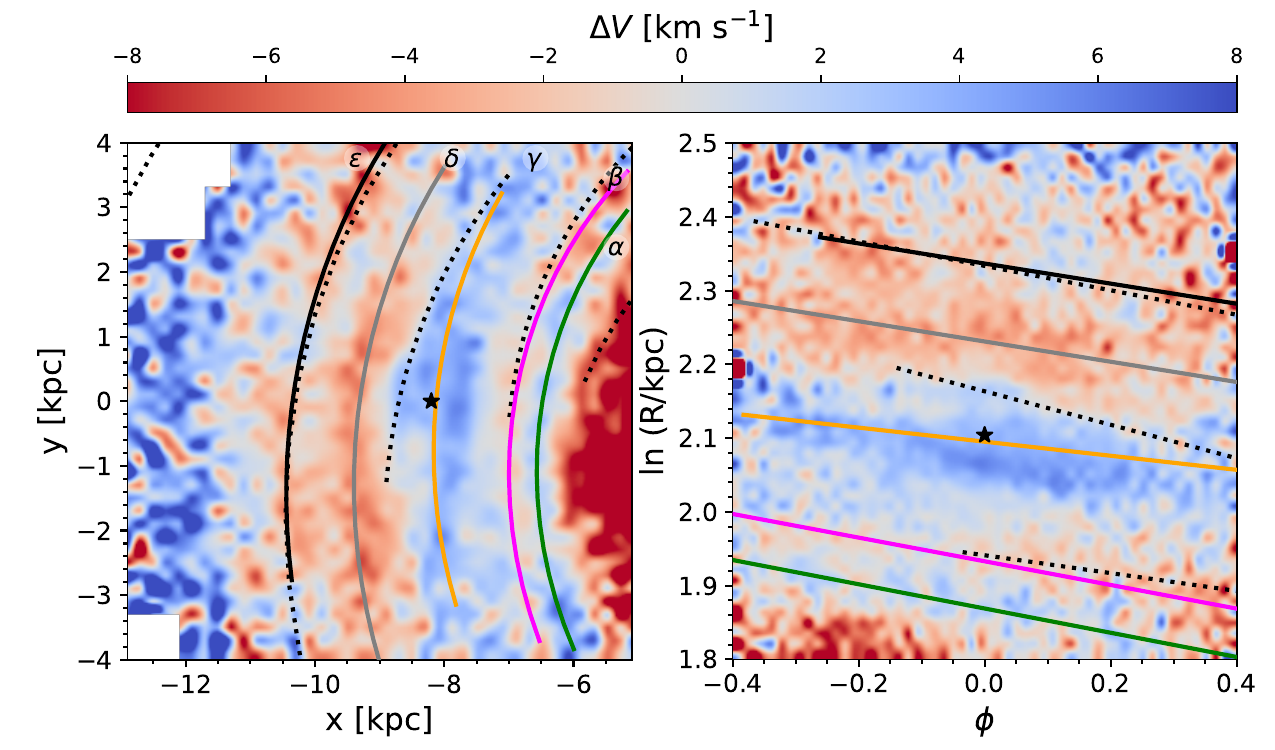}
\end{center}
\caption{Logarithmic spirals used to describe the structures in $\Delta V$ (colour solid lines). The dotted lines are the spiral arm models from \citet{Reid2014} (from right to left: Scutum-Centaurus, Sagittarius-Carina, Local, Perseus, and Outer arm).}
\label{fig:DV-fit}
\end{figure*}

To find the core of each $\Delta V$ kinematic spiral in the $x-y$ projection, we selected individually each structure as delimited by the change of sign in $\Delta V$; as for the frontier between $\delta$ and $\epsilon$, we used the ridge with the highest velocity. Note that a global shift in the absolute value of $\Delta V$ will result in a change on the width of the structures, but will hardly modify the core of each structure. We fitted the equation \ref{eq:spiral} using a non-linear least squares method to estimate the parameters $R_{\rm ref}$, and $\psi$. The values of the spiral fits are given in Table \ref{tab:fits}, and the resulting spiral curves are shown in Fig. \ref{fig:DV-fit} as solid colour lines (green for structure $\alpha$, magenta for $\beta$, orange for $\gamma$, grey for $\delta$, black for $\epsilon$).

\begin{table}
    {\centering
    \caption{Spiral kinematics fits}
    \begin{tabular}{c|c|ccl}
    \hline
    \hline
          &$R_{\rm ref}$ &$\psi$ & $\phi$ range&\\
 Spiral   & (kpc)    &   (deg)      &  (deg)  & Type\\
         \hline
           $\alpha$  & $6.48$   &  $9.34$     &  ($-32.6, \; 29.6$)  &  Inter-arm\\
           $\beta$   & $6.91$   &  $9.12$     &  ($-29.6, \; 34.5$)  &  Arm $^{\rm 1}$ \\
           $\gamma$  & $8.13$   &  $5.45$      & ($-21.9, \; 24.3$)  &  Inter-arm \\
           $\delta$  & $9.31$   &  $7.82$      & ($-24.6, \; 24.4$)  &  Arm $^{\rm 2}$   \\
           $\epsilon$ & $10.34$  &  $7.76$      & ($-14.9, \; 24.1$) &  Arm $^{\rm 3}$  \\
      \hline    
      \hline
    \end{tabular}}
    \\
  $^{\rm 1}$ Consistent with the Sagittarius-Carina arm from \citet{Reid2014}.\\
  $^{\rm 2}$ Arm-like structure without observational overdensity counterpart.\\ 
  $^{\rm 3}$ Consistent with the Perseus arm from \citet{Reid2014}.

    \label{tab:fits}
\end{table}

We can now compare these Gaia kinematic structures, presented in Fig. \ref{fig:DV}, with the MW spiral arms traced by young objects, presented by \citet{Reid2014}. In Fig. \ref{fig:DV-fit} we can see that the $\beta$ and $\epsilon$ structures, described by the magenta and black spirals, are in good agreement with the Sagittarius-Carina and Perseus arms, respectively.

\subsection{Galactic spiral arms in 3D}
\label{sec:3D}

By computing the contrast $\Delta V$ we found the regions of the Galactic plane where the rotation deviates from the axisymmetric velocity. Interestingly, these regions are arranged in spiral structures in the $x-y$ plane. Moreover, the full 3D information of the sample would allow us to explore $x-z$ and $y-z$ slices to have the complete picture of the kinematic spiral arms. 

To study the structure in 3D, with a good signal to noise ratio, we need to relax the constraint on $|z|$, while keeping all other constrains previously stated. In this section, we will use all the stars with $|z| < 1.0$ kpc; this increased z range, compared with sections \ref{sec:data} and \ref{sec:DV}, allows us to see the structure beyond the thin disc. With this new set of constraints, we end up with a subsample of 877,508 stars that belong to the thin and thick discs.

\begin{figure*}
\begin{center}
\includegraphics[width=\textwidth]{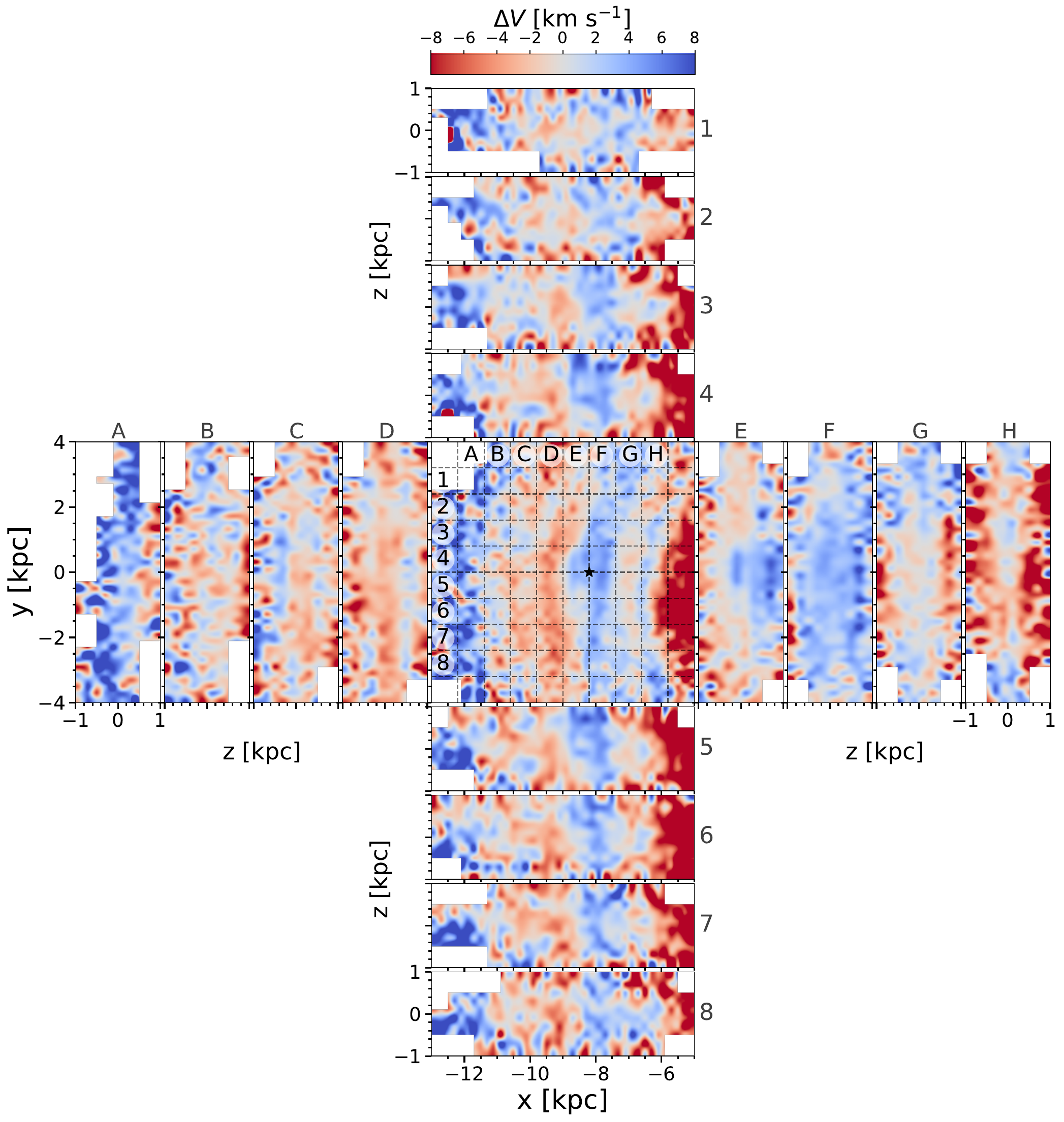}
\end{center}
\caption{Three-dimensional view of the residual rotational velocity $\Delta V$. Vertical panels are y-z projections from stars inside each of the vertical strips labelled from A to H in the central plot. Similarly, horizontal panels are x-z projections from the strips numbered from 1 to 8 in the central plot.}
\label{fig:V_xz}
\end{figure*}

In order to explore the 3D structure of the Galaxy, we divided the observed volume into rows and columns (labelled with numbers and letters, respectively), each with a width of 0.8 kpc. Fig. \ref{fig:V_xz} includes, at its centre, our original $\Delta V$ x-y map used in previous sections, as well as the column and row extracts. 

For the columns, we defined regions limited by a minimum and maximum value of x, while $-4\leq y\leq4$ and $-1\leq z\leq1$. Next, all stars inside each column are projected on the $y-z$ plane, and from here we apply the same procedure explained in sec. \ref{sec:DV}, i.e., divide this projection into cells coloured by the local value of $\Delta V$. The resulting projections are shown in the vertical subplots of Fig. \ref{fig:V_xz}, and labelled from A to H.

Similarly, for the rows, we delimit the maximum and minimum values of y, while $-13\leq x\leq-5$ and $-1\leq z\leq1$. Then, all stars inside each of these horizontal regions are projected on the x-z plane. The resulting projections are shown in the horizontal subplots of Fig. \ref{fig:V_xz}, and numbered from 1 to 8.

It is worth mentioning that for the computation of $\Delta V$ (equation \ref{eq:DV}) we are using the same averaged rotation curve $V_{\it avg}(R)$ used in previous sections to represent stars higher above the plane, even when we expect the rotational velocity to diminish as we exit the plane. We do this, mainly, for the following three reasons: i) a family of universal curves fit to each relevant height would have poor signal to noise and would lack physical meaning; ii) we expect such simplification to result in a slowing down (reddening) of the bins away from the plane, this in turn will be a competing effect with the speeding up (blue) we expect from stars that are above/below the arms, once they are not longer dominated by the dynamics of the spiral structure, thus we should be able to distinguish between these two effects; and, finally iii) it is beyond the scope of this paper to develop a 3D extension of the universal curve.

By looking at the kinematic trace of the Perseus arm, 
present in panel C, we find a change in the velocity signal going from negative (red) at small heights $-200 <z< 400$ pc to positive (blue) at heights slightly above such range; from this change in behaviour, we find that the spiral arm vertical thickness amounts to $\sim600$ pc (full height). Note that, for heights above $|z|>600$ pc, the stars slow down again (red); this is due to the diminishing rotation velocity away from the plane, that becomes evident at such heights. 

Similarly, by looking at the Sagittarius-Carina arm, present in panel G, we find its vertical thickness to be $\sim400$ pc (full height). This is the first determination of the thickness of the kinematic spiral arms; it has only been possible with the quality of the EDR3 Gaia data.

Also notice that panels 4 and 5 show the perpendicular projection $x-z$ that captures the thickness of Perseus and Sagittarius-Carina at the same time. Here, we notice that the kinematic counterpart of the Perseus arm is roughly symmetric around the $z=0$ plane; while that of Sagittarius-Carina arm is not, it is clear that its amplitude in $\Delta V$ is larger below the Galactic plane.

It should be noted that both these thicknesses depend on the exact value of the assumed axisymmetric rotation. Our particular choice for $V_{avg} (R)$ represents a good fit to the the available data; thus we are confident on the approximate thicknesses obtained in this section.

\subsection{Radial Velocity}
\label{sec:radial}

In this section we look for kinematic structures traced by the radial velocities, $U$, of our stellar sample. As in Sec. \ref{sec:DV}, we project the stars in our sample on the x-y plane which we divide into cells. Then, each cell is coloured by the median value of $U$ from all stars in the cell.

\begin{figure*}
\begin{center}
\includegraphics[width=\textwidth]{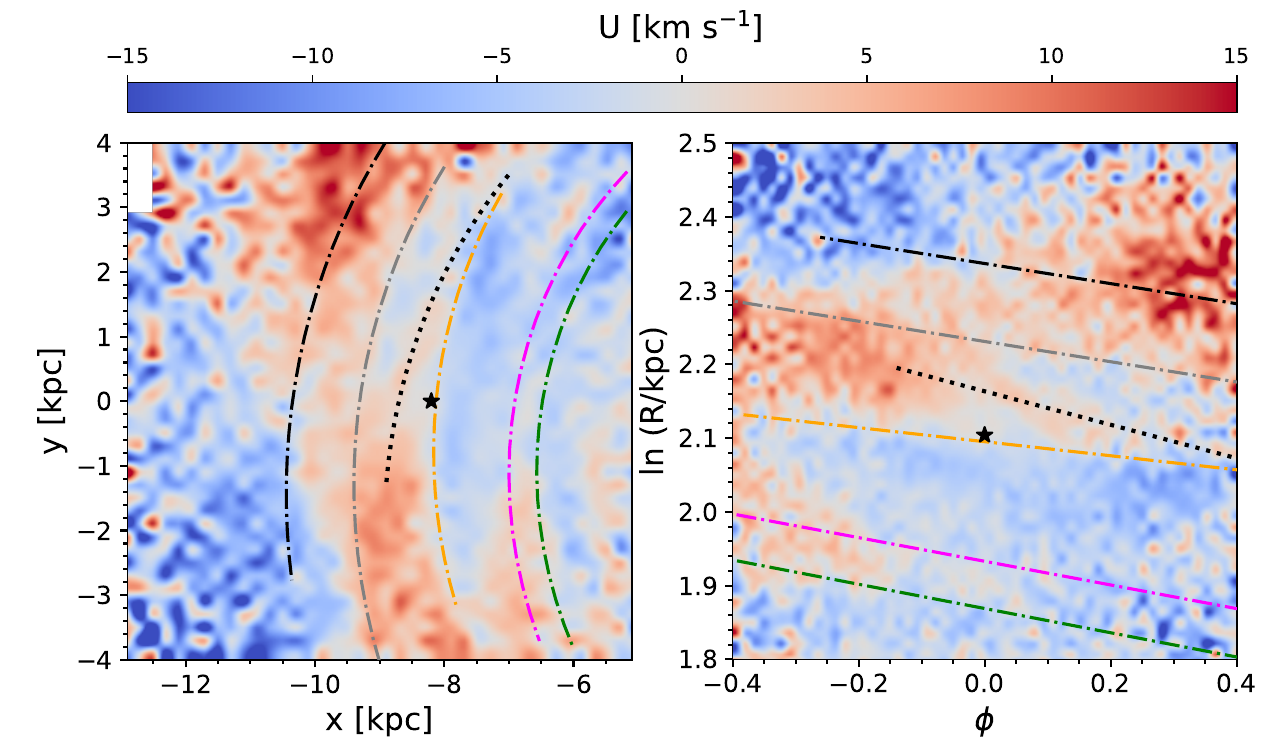}
\end{center}
\caption{Kinematic structure in the stellar $U$ velocity projected on the Galactic plane. For comparison we include the logarithmic spirals (dash-dotted lines) used to fit the $\Delta V$ structures (Fig. \ref{fig:DV-fit}). The dotted line is the Local arm from \citet{Reid2014}. }
\label{fig:U}
\end{figure*}

Fig. \ref{fig:U}, left panel, shows the distribution of radial velocities projected on the Galactic mid-plane. We see that this plane is also rich in substructures, although they do not have the same shapes as the ones found with $\Delta V$, nor resemble clear spiral bands. In the right panel ($\phi-ln(R)$ plane) it can be seen that some of these structures resemble elongated triangles and correspond to some form of spiral triangle in the left panel (physical plane).

To analyse whether the structures in $U$ have a connection with the Galactic spiral arms, in Fig. \ref{fig:U} we overplot the logarithmic spirals used to fit the $\Delta V$ structures (Fig. \ref{fig:DV-fit}). In general, the structures in $U$ cannot be entirely described with the logarithmic spirals used for the $\Delta V$ structures. However, we notice some similarities: i) the upper half of the black curve, that matches well with the Perseus arm, resembles the mid region of the large red structure in the upper part of the plot; ii) the lower segment of the magenta curve, that follows the Sagittarius-Carina arm, crosses through a red $U$ structure in the bottom right of the plot.; iii) the grey spiral has no clear relation with any red or blue structure, residing near the frontier between blue and red, but changing sides near the middle of our plot; iv) as for the green and orange spirals, that in $\Delta V$ trace the central part of the inter-arms, in Fig. \ref{fig:U} they are closer to interfaces where $U$ changes from positive to negative values.

Nonetheless, in Fig. \ref{fig:U} we found an extra structure; close to the Sun's position is a red ($U>0$) diagonal spiral fragment that follows the location of the Local arm from \citet{Reid2014}, plotted here with a dotted line. When looking at the right panel, it is apparent that this structure ends at phi=0.25, however it is not clear whether the beginning lies close to $\phi=-0.05$ or it extends from $\phi<-0.4$.

Recently, \citet{Eilers2020} found a spiral signature using the mean Galactocentric radial velocities of red giant stars in the mid-plane. By comparing with a non-axisymmetric potential model, they identified the interface between positive (on the inside) and negative (on the outside) velocity values as the location of spiral arms density peaks. We cannot find such clear spiral signature in $U$; however, our analysis is centred on a smaller region of the Galactic disc.

\subsection{Vertical Velocity}
\label{sec:vertical}

Across an axisymmetric stellar disc the mean vertical velocity would be zero. However, in the presence of additional vertically symmetric components (such as idealised arms or bars), some local deviations are expected to happen, these deviations would also be symmetrical across the plane, and would theoretically be zero when projected on the plane; however, the presence of dust clouds above or below the plane can produce a bias, and thus some signal could be expected. In a real galaxy, additional perturbations (both external and internal) are expected. Further discussion on the perturbations in the vertical velocities of the disc, either in the stellar or gas component, can be found in the literature \citep{Weinberg1991,Faure2014,Monari2016a,Alves2020,LopezCorredoira2020}. In this section we analyse the vertical velocity W of our sample. 

\begin{figure*}
\begin{center}
\includegraphics[width=\textwidth]{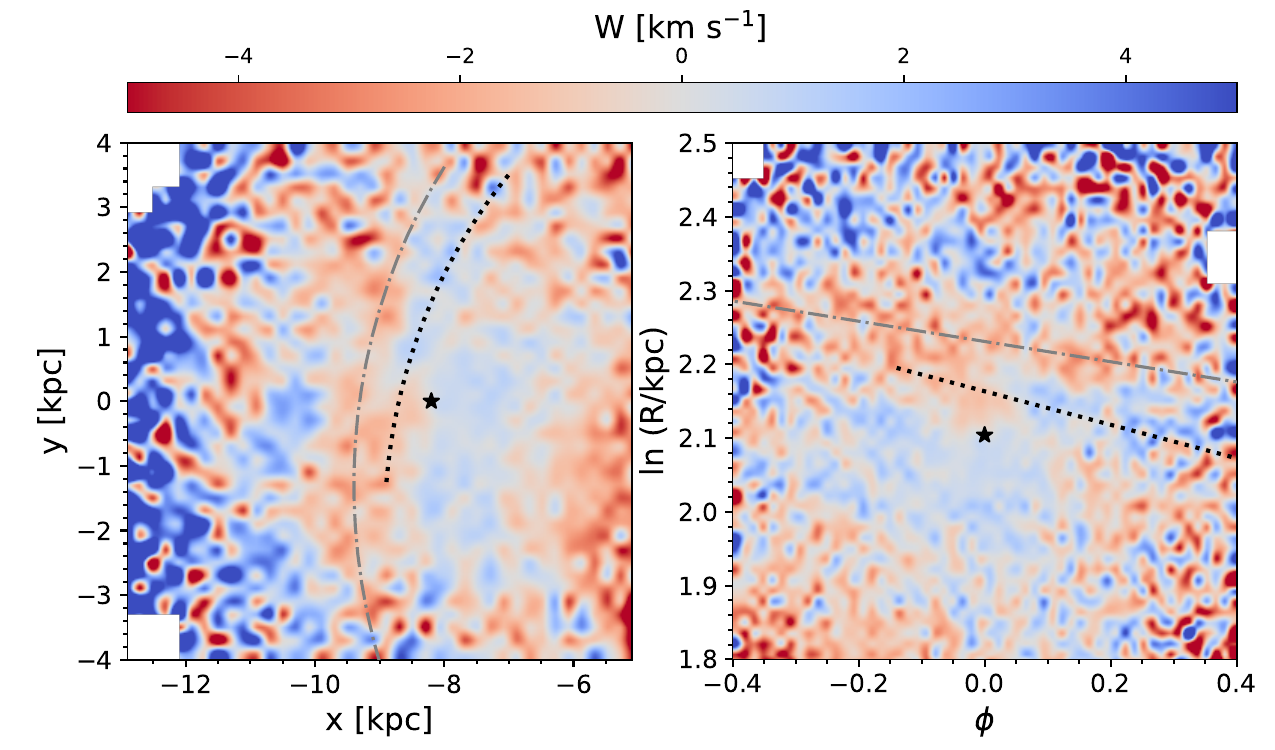}
\end{center}
\caption{Footprint in the $W$ component of the stellar velocity projected on the Galactic plane. The dash-dotted line is the spiral curve used to describe the $\delta$ structure in $\Delta V$ (Fig. \ref{fig:DV-fit}), included here for comparison. The dotted line is the Local arm from \citet{Reid2014}.}
\label{fig:W}
\end{figure*}

Fig \ref{fig:W} shows the x-y plane coloured by the local $W$ values; there we find extended regions with non zero mean vertical velocity. Of particular interest is the large vertical red region (negative W values) in the middle of the graph, with a mean vertical velocity of $W\approx-2$ km/s. It looks similar to the spiral $\delta$ found in $\Delta V$ at the same position, just behind of the Perseus arm. We confirm this correlation by overplotting the grey spiral obtained in Fig. \ref{fig:DV-fit}. This result is not unusual; \citet{Eilers2020}, using red giant stars from APOGEE DR14, found structures in the vertical velocity map (their figure A1) resembling those found in the radial velocity map. 

Although the nature of the perturber causing these vertical motions is not clear, some inferences can be made: 
i) This feature seems to be related to the $\delta$ spiral structure found in section \ref{sec:DV}. 
ii) It cannot be the effect of molecular clouds obscuring one side of the plane. 
iii) Its wide area suggests that this feature is not produced by the gravitational effect of giant molecular clouds (or any other local perturbation) that lie off the disc plane.
iv) A plausible scenario is that it originated from an extragalactic perturbation \citep[perhaps a dwarf galaxy crossing the Galactic plane; e.g.][]{Laporte2019}. 
v) If the perturber is indeed extragalactic in nature, the size of the feature suggests that this perturbation has had time to lengthen, due to differential rotation; implying that the triggering event was not recent.
vi) Moreover, its observed shape and pitch angle are very similar to those of Perseus arm; on top of that, this feature lags behind Perseus arm, suggesting a connection between this structure with both, Perseus arm and the Galactic rotation.

There is an additional faint blue structure with a Galactocentric distance of $R\sim$8.3 kpc and velocity $W\approx1$ km s$^{-1}$. It seems to be connected to the local arm; this perturbation could also be related to the Radcliffe gas wave recently observed by \citet{Alves2020}. Whether the original perturbation is somehow related to the formation mechanism of the local arm, or it is an independent perturbation propagated by the local arm, is beyond the scope of this paper.

\section{Discussion and Conclusions}
\label{discussion}

We have analysed the three velocity components of thin disc stars from Gaia EDR3. The projections on the mid-plane coloured by each of the three stellar velocity components are rich in substructures; of particular interest are the local deviations from axisymmetric rotation, that reveal detailed large-scale kinematic spiral structures. All this work has been possible due to the large breath and exquisite quality available from the Gaia mission.

A big advantage of our approach is that the inherent biases in star counts due to Galactic extinction do not interfere with our determinations, this is because we are not measuring star number densities (at most these issues in star counts would decrease our s/n but without introducing any bias). The signal we find depends on the response of hundreds of thousands of stars (both young and old) to the local mass distribution of the different Galactic components; unlike the signal that can be detected using molecular clouds, \ion{H}{ii} regions, or other young objects, that often depends on small number statistics.

Regarding the dependence of our results on the specific form of $V_{\it avg}(R)$: the data is insufficient to properly determine $V_{\it avg}(R)$ and the universal rotation curve model, adopted in this work, has many limitations; however, If we used a different functional form for $V_{\it avg}(R)$, the observed kinematic spiral structure would have the same pitch angle. In one dimension the intersection between the average data points and the RC model would change, but the maxima and minima would remain; in two dimensions, this would manifest as different contours to the structures present in such range, yet the locii of the maxima and minima of each structure would be nearly unaffected.

Based on stellar dynamics, it is expected that stars in spiral arms rotate slower than the averaged azimuthal velocity, while stars in the inter-arms rotate faster. We studied the residual rotation velocity, $\Delta V$, and found five structures: three of our structures are slow (red) and  we associated them with spiral arms (named $\beta$, $\delta$, and $\epsilon$); while the other two are fast (blue) and we associated them with inter-arms (named $\alpha$ and $\gamma$).

Comparing our structures with those by \citet{Reid2014}, we found that the location of the $\beta$ and $\epsilon$ spirals are in good agreement with the Sagittarius-Carina arm (dominated by gas) and the Perseus arm (more massive, with a backbone made of stars), respectively. Our best fits with logarithmic spirals give pitch angles of 9.12$\deg$ and 7.76$\deg$ for our Sagittarius-Carina and Perseus counterparts, respectively. The kinematic differences between the arms and inter-arms are in favour of MW spiral arms that do not corotate with the disc; i.e., since corotating arms require for arm stars, inter-arm stars as well as the stellar disc to have the same rotation speed at a given radius \citep[e.g.][]{Kawata2014}, neither the $\beta$ nor the $\epsilon$ spirals are corotating with the disc.

Since we are measuring the locus of the kinematic signal instead of the spatial location of stellar density peaks, these two signals are not necessarily expected to coincide, as the relation between density and kinematics is not straight forward. Interestingly, for two of our kinematic arms, we do find location and pitch angle coincidence with previous determinations \citep{Reid2014}, the deviations from their known overdensity counterparts are minimal, consistent within $1-\sigma$ uncertainties. 
 
Besides $\beta$ and $\epsilon$, the last kinematic spiral arm, $\delta$, does not appear to have an overdensity counterpart, neither in stars nor in gas. However, this structure is close to the Perseus arm, and both share similar pitch angles. As Perseus is one of the two massive stellar arms, strong enough to trigger a wide kinematic response as it sweeps the disc, suggesting the $\delta$ structure could represent an out of phase response to Perseus arm ($\epsilon$).

It is worth to mention that each kinematic spiral found in Fig. \ref{fig:DV} can be described by a logarithmic spiral (Fig. \ref{fig:DV-fit}). However, these structures are not continuous, they have several gaps, and are made up by clumps, revealing a flocculent nature in the kinematics of the Galactic spiral arms. This kinematic flocculence seems to be a counterpart of the known density flocculence.

The last action with the $\Delta V$ component was to generate x-z and y-z maps. The result is a set of vertical slices of the disc (Fig. \ref{fig:V_xz}), that reveal the thickness of the spiral arms. We found that the kinematic counterpart of Perseus arm has a vertical thickness of $\sim$600 pc, while that of Sagittarius-Carina arm has a thickness of $\sim$400 pc.

We have also studied the $U$ velocity. We presented a similar x-y map, but coloured with the local median values of the radial velocity $U$ (Fig. \ref{fig:U}). The signal is quite different between $U$ and $\Delta V$: since we do not find spiral bands equivalent to either arms or inter-arms, instead finding elongated triangular features that cannot be identified as spirals.

Finally, regarding the vertical component $W$, we also analyse this velocity in the x-y projection (Fig.\ref{fig:W}). The simplest models to study the vertical velocities, $W$, would expect this map to be nothing more than noise. However, we do find two structures: the clearest one is a red arch at $R \sim 9.5$ kpc akin to the $\delta$ structure; the second one is a faint blue arch at $R \sim 8.2$ kpc akin to the Local arm, but only present ahead of the Sun (i.e. y$>0$). 

The brightest structure found in $W$ is consistent with the the spiral arm $\delta$ found in Fig. \ref{fig:DV}; the presence and shape of this structure in the vertical velocity component suggest a possible scenario of a dwarf galaxy crossing the plane in the not-too-recent past. When including all three velocity components, this feature presents significant signal in both $\Delta V$ and $W$, as well as a faint signal in $U$; even if there is no known significant signature of an overdensity in any type of objects, the $\delta$ structure is a complex kinematic perturbation that resides within the Galactic disc. 

The local arm structure seems to be present in two of the velocity components, $U$ and $W$; but $\Delta V$ seems to be less sensitive to the local arm as it is to either the Sagittarius-Carina and Perseus arms, which are clearly traced in Fig. \ref{fig:DV-fit}. This is likely due to the nature of the Local arm, which since early studies has been considered a fragment or spur \citep{vandeHulst1954,Georgelin1976}. Its density contrast, either in gas or stars, is not as large as in any of the four major arms; evidence of this can be seen in the RC of Fig. \ref{fig:fits}; while Sagittarius-Carina imprints a local minimum in the RC at $\sim$7 kpc, and Perseus does the same at $\sim$10.7 kpc, the Local arm is not massive enough to modify the RC in such a clear way. However, the components $U$ and $W$ of the velocity appear to be sensitive enough to trace the Local arm. Figures \ref{fig:U} and \ref{fig:W} reveal a diagonal structure close to the Sun's position, that falls under the overplotted Local arm (dotted line) as measured by \citet{Reid2014}.

Overall, our approach has allowed us to uncover six kinematic structures, three of which represent the first time the kinematic counterparts of Perseus, Sagittarius-Carina, and the Local arm are observed; in particular, using kinematics, this is the first time the vertical heights of Perseus and Sagittarius-Carina are measured. The other three represent structures not studied before: the kinematic inter-arms we call $\alpha$ and $\gamma$, but most notably $\delta$, which represents a new kind of kinematic arm not observed before. All this work was only possible thanks to the high-quality data from Gaia EDR3. We are convinced that with future updates from the Gaia Collaboration all the structures found here should improve and more of them will be detected.

\section*{Acknowledgements}

We thank the referee for a critical reading of the manuscript and several useful suggestions. We also acknowledge the support from DGAPA-PAPIIT IA101520, IA103122, IG100319, and IG100622 grants. We acknowledge DGTIC-UNAM for providing HPC resources on the Cluster Supercomputer Miztli.  
 
\section*{Data availability}

The data that support the findings of this study are available from the corresponding author, upon reasonable request.

\bibliographystyle{mnras}
\bibliography{ref}

\begin{thebibliography}{}
\makeatletter
\relax
\def\mn@urlcharsother{\let\do\@makeother \do\$\do\&\do\#\do\^\do\_\do\%\do\~}
\def\mn@doi{\begingroup\mn@urlcharsother \@ifnextchar [ {\mn@doi@}
  {\mn@doi@[]}}
\def\mn@doi@[#1]#2{\def\@tempa{#1}\ifx\@tempa\@empty \href
  {http://dx.doi.org/#2} {doi:#2}\else \href {http://dx.doi.org/#2} {#1}\fi
  \endgroup}
\def\mn@eprint#1#2{\mn@eprint@#1:#2::\@nil}
\def\mn@eprint@arXiv#1{\href {http://arxiv.org/abs/#1} {{\tt arXiv:#1}}}
\def\mn@eprint@dblp#1{\href {http://dblp.uni-trier.de/rec/bibtex/#1.xml}
  {dblp:#1}}
\def\mn@eprint@#1:#2:#3:#4\@nil{\def\@tempa {#1}\def\@tempb {#2}\def\@tempc
  {#3}\ifx \@tempc \@empty \let \@tempc \@tempb \let \@tempb \@tempa \fi \ifx
  \@tempb \@empty \def\@tempb {arXiv}\fi \@ifundefined
  {mn@eprint@\@tempb}{\@tempb:\@tempc}{\expandafter \expandafter \csname
  mn@eprint@\@tempb\endcsname \expandafter{\@tempc}}}

\bibitem[\protect\citeauthoryear{{Alves} et~al.,}{{Alves}
  et~al.}{2020}]{Alves2020}
{Alves} J.,  et~al., 2020, \mn@doi [\nat] {10.1038/s41586-019-1874-z}, \href
  {https://ui.adsabs.harvard.edu/abs/2020Natur.578..237A} {578, 237}

\bibitem[\protect\citeauthoryear{{Antoja}, {Figueras}, {Romero-G{\'o}mez},
  {Pichardo}, {Valenzuela}  \& {Moreno}}{{Antoja} et~al.}{2011}]{Antoja2011}
{Antoja} T.,  {Figueras} F.,  {Romero-G{\'o}mez} M.,  {Pichardo} B.,
  {Valenzuela} O.,   {Moreno} E.,  2011, \mn@doi [\mnras]
  {10.1111/j.1365-2966.2011.19190.x}, \href
  {https://ui.adsabs.harvard.edu/abs/2011MNRAS.418.1423A} {418, 1423}

\bibitem[\protect\citeauthoryear{{Antoja} et~al.,}{{Antoja}
  et~al.}{2018}]{Antoja2018}
{Antoja} T.,  et~al., 2018, \mn@doi [\nat] {10.1038/s41586-018-0510-7}, \href
  {https://ui.adsabs.harvard.edu/abs/2018Natur.561..360A} {561, 360}

\bibitem[\protect\citeauthoryear{{Baba}, {Morokuma-Matsui}, {Miyamoto}, {Egusa}
   \& {Kuno}}{{Baba} et~al.}{2016}]{Baba2016}
{Baba} J.,  {Morokuma-Matsui} K.,  {Miyamoto} Y.,  {Egusa} F.,   {Kuno} N.,
  2016, \mn@doi [\mnras] {10.1093/mnras/stw987}, \href
  {https://ui.adsabs.harvard.edu/abs/2016MNRAS.460.2472B} {460, 2472}

\bibitem[\protect\citeauthoryear{{Barros}, {P{\'e}rez-Villegas}, {L{\'e}pine},
  {Michtchenko}  \& {Vieira}}{{Barros} et~al.}{2020}]{Barros2020}
{Barros} D.~A.,  {P{\'e}rez-Villegas} A.,  {L{\'e}pine} J. R.~D.,
  {Michtchenko} T.~A.,   {Vieira} R. S.~S.,  2020, \mn@doi [\apj]
  {10.3847/1538-4357/ab59d1}, \href
  {https://ui.adsabs.harvard.edu/abs/2020ApJ...888...75B} {888, 75}

\bibitem[\protect\citeauthoryear{{Becker}}{{Becker}}{1963}]{Becker1963}
{Becker} W.,  1963, \zap, \href
  {https://ui.adsabs.harvard.edu/abs/1963ZA.....57..117B} {57, 117}

\bibitem[\protect\citeauthoryear{{Becker} \& {Fenkart}}{{Becker} \&
  {Fenkart}}{1970}]{Becker1970}
{Becker} W.,  {Fenkart} R.~B.,  1970, in {Becker} W.,  {Kontopoulos} G.~I.,
  eds,  Vol. 38, The Spiral Structure of our Galaxy. p.~205

\bibitem[\protect\citeauthoryear{{Bennett} \& {Bovy}}{{Bennett} \&
  {Bovy}}{2019}]{Bennett2019}
{Bennett} M.,  {Bovy} J.,  2019, \mn@doi [\mnras] {10.1093/mnras/sty2813},
  \href {https://ui.adsabs.harvard.edu/abs/2019MNRAS.482.1417B} {482, 1417}

\bibitem[\protect\citeauthoryear{{Binney}, {Gerhard}  \& {Spergel}}{{Binney}
  et~al.}{1997}]{Binney1997}
{Binney} J.,  {Gerhard} O.,   {Spergel} D.,  1997, \mn@doi [\mnras]
  {10.1093/mnras/288.2.365}, \href
  {https://ui.adsabs.harvard.edu/abs/1997MNRAS.288..365B} {288, 365}

\bibitem[\protect\citeauthoryear{{Bland-Hawthorn} \&
  {Gerhard}}{{Bland-Hawthorn} \& {Gerhard}}{2016}]{Bland-Hawthorn2016}
{Bland-Hawthorn} J.,  {Gerhard} O.,  2016, \mn@doi [\araa]
  {10.1146/annurev-astro-081915-023441}, \href
  {https://ui.adsabs.harvard.edu/abs/2016ARA&A..54..529B} {54, 529}

\bibitem[\protect\citeauthoryear{{Bovy}}{{Bovy}}{2015}]{Bovy2015}
{Bovy} J.,  2015, \mn@doi [\apjs] {10.1088/0067-0049/216/2/29}, \href
  {https://ui.adsabs.harvard.edu/abs/2015ApJS..216...29B} {216, 29}

\bibitem[\protect\citeauthoryear{{Buta}}{{Buta}}{1989}]{Buta1989}
{Buta} R.,  1989, {Galaxy Morphology}.
p.~151

\bibitem[\protect\citeauthoryear{{Castro-Ginard} et~al.,}{{Castro-Ginard}
  et~al.}{2021}]{Castro-Ginard2021}
{Castro-Ginard} A.,  et~al., 2021, arXiv e-prints, \href
  {https://ui.adsabs.harvard.edu/abs/2021arXiv210504590C} {p. arXiv:2105.04590}

\bibitem[\protect\citeauthoryear{{D'Onghia}, {Vogelsberger}  \&
  {Hernquist}}{{D'Onghia} et~al.}{2013}]{DOnghia2013}
{D'Onghia} E.,  {Vogelsberger} M.,   {Hernquist} L.,  2013, \mn@doi [\apj]
  {10.1088/0004-637X/766/1/34}, \href
  {https://ui.adsabs.harvard.edu/abs/2013ApJ...766...34D} {766, 34}

\bibitem[\protect\citeauthoryear{{Dias} \& {L{\'e}pine}}{{Dias} \&
  {L{\'e}pine}}{2005}]{Dias2005}
{Dias} W.~S.,  {L{\'e}pine} J.~R.~D.,  2005, \mn@doi [\apj] {10.1086/431456},
  \href {https://ui.adsabs.harvard.edu/abs/2005ApJ...629..825D} {629, 825}

\bibitem[\protect\citeauthoryear{{Drimmel} \& {Spergel}}{{Drimmel} \&
  {Spergel}}{2001}]{Drimmel2001}
{Drimmel} R.,  {Spergel} D.~N.,  2001, \mn@doi [\apj] {10.1086/321556}, \href
  {https://ui.adsabs.harvard.edu/abs/2001ApJ...556..181D} {556, 181}

\bibitem[\protect\citeauthoryear{{Eilers}, {Hogg}, {Rix}, {Frankel}, {Hunt},
  {Fouvry}  \& {Buck}}{{Eilers} et~al.}{2020}]{Eilers2020}
{Eilers} A.-C.,  {Hogg} D.~W.,  {Rix} H.-W.,  {Frankel} N.,  {Hunt} J. A.~S.,
  {Fouvry} J.-B.,   {Buck} T.,  2020, \mn@doi [\apj]
  {10.3847/1538-4357/abac0b}, \href
  {https://ui.adsabs.harvard.edu/abs/2020ApJ...900..186E} {900, 186}

\bibitem[\protect\citeauthoryear{{Elmegreen}}{{Elmegreen}}{1990}]{Elmegreen1990}
{Elmegreen} B.~G.,  1990, \mn@doi [Annals of the New York Academy of Sciences]
  {10.1111/j.1749-6632.1990.tb27410.x}, \href
  {https://ui.adsabs.harvard.edu/abs/1990NYASA.596...40E} {596, 40}

\bibitem[\protect\citeauthoryear{{Elmegreen} \& {Elmegreen}}{{Elmegreen} \&
  {Elmegreen}}{1984}]{Elmegreen1984}
{Elmegreen} D.~M.,  {Elmegreen} B.~G.,  1984, \mn@doi [\apjs] {10.1086/190922},
  \href {https://ui.adsabs.harvard.edu/abs/1984ApJS...54..127E} {54, 127}

\bibitem[\protect\citeauthoryear{{Elmegreen} \& {Elmegreen}}{{Elmegreen} \&
  {Elmegreen}}{1986}]{Elmegreen1986}
{Elmegreen} B.~G.,  {Elmegreen} D.~M.,  1986, \mn@doi [\apj] {10.1086/164795},
  \href {https://ui.adsabs.harvard.edu/abs/1986ApJ...311..554E} {311, 554}

\bibitem[\protect\citeauthoryear{{Englmaier} \& {Gerhard}}{{Englmaier} \&
  {Gerhard}}{1999}]{Englmaier1999}
{Englmaier} P.,  {Gerhard} O.,  1999, \mn@doi [\mnras]
  {10.1046/j.1365-8711.1999.02280.x}, \href
  {https://ui.adsabs.harvard.edu/abs/1999MNRAS.304..512E} {304, 512}

\bibitem[\protect\citeauthoryear{{Faure}, {Siebert}  \& {Famaey}}{{Faure}
  et~al.}{2014}]{Faure2014}
{Faure} C.,  {Siebert} A.,   {Famaey} B.,  2014, \mn@doi [\mnras]
  {10.1093/mnras/stu428}, \href
  {https://ui.adsabs.harvard.edu/abs/2014MNRAS.440.2564F} {440, 2564}

\bibitem[\protect\citeauthoryear{{Fenkart} \& {Binggeli}}{{Fenkart} \&
  {Binggeli}}{1979}]{Fenkart1979}
{Fenkart} R.~P.,  {Binggeli} B.,  1979, \aaps, \href
  {https://ui.adsabs.harvard.edu/abs/1979A&AS...35..271F} {35, 271}

\bibitem[\protect\citeauthoryear{{Foreman-Mackey}, {Hogg}, {Lang}  \&
  {Goodman}}{{Foreman-Mackey} et~al.}{2013}]{Foreman-Mackey2013}
{Foreman-Mackey} D.,  {Hogg} D.~W.,  {Lang} D.,   {Goodman} J.,  2013, \mn@doi
  [\pasp] {10.1086/670067}, \href
  {https://ui.adsabs.harvard.edu/abs/2013PASP..125..306F} {125, 306}

\bibitem[\protect\citeauthoryear{{Gaia Collaboration} et~al.,}{{Gaia
  Collaboration} et~al.}{2018}]{Gaia2018}
{Gaia Collaboration} et~al., 2018, \mn@doi [\aap]
  {10.1051/0004-6361/201833051}, \href
  {https://ui.adsabs.harvard.edu/abs/2018A&A...616A...1G} {616, A1}

\bibitem[\protect\citeauthoryear{{Gaia Collaboration} et~al.,}{{Gaia
  Collaboration} et~al.}{2021}]{Gaia2021}
{Gaia Collaboration} et~al., 2021, \mn@doi [\aap]
  {10.1051/0004-6361/202039657}, \href
  {https://ui.adsabs.harvard.edu/abs/2021A&A...649A...1G} {649, A1}

\bibitem[\protect\citeauthoryear{{Georgelin} \& {Georgelin}}{{Georgelin} \&
  {Georgelin}}{1976}]{Georgelin1976}
{Georgelin} Y.~M.,  {Georgelin} Y.~P.,  1976, \aap, \href
  {https://ui.adsabs.harvard.edu/abs/1976A&A....49...57G} {49, 57}

\bibitem[\protect\citeauthoryear{{Grand}, {Kawata}  \& {Cropper}}{{Grand}
  et~al.}{2012}]{Grand2012}
{Grand} R. J.~J.,  {Kawata} D.,   {Cropper} M.,  2012, \mn@doi [\mnras]
  {10.1111/j.1365-2966.2012.20411.x}, \href
  {https://ui.adsabs.harvard.edu/abs/2012MNRAS.421.1529G} {421, 1529}

\bibitem[\protect\citeauthoryear{{Gravity Collaboration} et~al.,}{{Gravity
  Collaboration} et~al.}{2019}]{Gravity2019}
{Gravity Collaboration} et~al., 2019, \mn@doi [\aap]
  {10.1051/0004-6361/201935656}, \href
  {https://ui.adsabs.harvard.edu/abs/2019A&A...625L..10G} {625, L10}

\bibitem[\protect\citeauthoryear{{Hayden} et~al.,}{{Hayden}
  et~al.}{2015}]{Hayden2015}
{Hayden} M.~R.,  et~al., 2015, \mn@doi [\apj] {10.1088/0004-637X/808/2/132},
  \href {https://ui.adsabs.harvard.edu/abs/2015ApJ...808..132H} {808, 132}

\bibitem[\protect\citeauthoryear{{Hou} \& {Han}}{{Hou} \&
  {Han}}{2014}]{Hou2014}
{Hou} L.~G.,  {Han} J.~L.,  2014, \mn@doi [\aap] {10.1051/0004-6361/201424039},
  \href {https://ui.adsabs.harvard.edu/abs/2014A&A...569A.125H} {569, A125}

\bibitem[\protect\citeauthoryear{{Hunt}, {Hong}, {Bovy}, {Kawata}  \&
  {Grand}}{{Hunt} et~al.}{2018}]{Hunt2018}
{Hunt} J. A.~S.,  {Hong} J.,  {Bovy} J.,  {Kawata} D.,   {Grand} R. J.~J.,
  2018, \mn@doi [\mnras] {10.1093/mnras/sty2532}, \href
  {https://ui.adsabs.harvard.edu/abs/2018MNRAS.481.3794H} {481, 3794}

\bibitem[\protect\citeauthoryear{{Kalnajs}}{{Kalnajs}}{1973}]{Kalnajs1973}
{Kalnajs} A.~J.,  1973, \mn@doi [\pasa] {10.1017/S1323358000013461}, \href
  {https://ui.adsabs.harvard.edu/abs/1973PASA....2..174K} {2, 174}

\bibitem[\protect\citeauthoryear{{Katz} et~al.,}{{Katz}
  et~al.}{2019}]{Katz2019}
{Katz} D.,  et~al., 2019, \mn@doi [\aap] {10.1051/0004-6361/201833273}, \href
  {https://ui.adsabs.harvard.edu/abs/2019A&A...622A.205K} {622, A205}

\bibitem[\protect\citeauthoryear{{Kawata}, {Hunt}, {Grand}, {Pasetto}  \&
  {Cropper}}{{Kawata} et~al.}{2014}]{Kawata2014}
{Kawata} D.,  {Hunt} J. A.~S.,  {Grand} R. J.~J.,  {Pasetto} S.,   {Cropper}
  M.,  2014, \mn@doi [\mnras] {10.1093/mnras/stu1292}, \href
  {https://ui.adsabs.harvard.edu/abs/2014MNRAS.443.2757K} {443, 2757}

\bibitem[\protect\citeauthoryear{{Laporte}, {Minchev}, {Johnston}  \&
  {G{\'o}mez}}{{Laporte} et~al.}{2019}]{Laporte2019}
{Laporte} C. F.~P.,  {Minchev} I.,  {Johnston} K.~V.,   {G{\'o}mez} F.~A.,
  2019, \mn@doi [\mnras] {10.1093/mnras/stz583}, \href
  {https://ui.adsabs.harvard.edu/abs/2019MNRAS.485.3134L} {485, 3134}

\bibitem[\protect\citeauthoryear{{Lin} \& {Shu}}{{Lin} \&
  {Shu}}{1964}]{Lin1964}
{Lin} C.~C.,  {Shu} F.~H.,  1964, \mn@doi [\apj] {10.1086/147955}, \href
  {https://ui.adsabs.harvard.edu/abs/1964ApJ...140..646L} {140, 646}

\bibitem[\protect\citeauthoryear{{Lindegren} et~al.,}{{Lindegren}
  et~al.}{2021}]{Lindegren2021}
{Lindegren} L.,  et~al., 2021, \mn@doi [\aap] {10.1051/0004-6361/202039653},
  \href {https://ui.adsabs.harvard.edu/abs/2021A&A...649A...4L} {649, A4}

\bibitem[\protect\citeauthoryear{{L{\'o}pez-Corredoira}, {Garz{\'o}n}, {Wang},
  {Sylos Labini}, {Nagy}, {Chrob{\'a}kov{\'a}}, {Chang}  \&
  {Villarroel}}{{L{\'o}pez-Corredoira} et~al.}{2020}]{LopezCorredoira2020}
{L{\'o}pez-Corredoira} M.,  {Garz{\'o}n} F.,  {Wang} H.~F.,  {Sylos Labini} F.,
   {Nagy} R.,  {Chrob{\'a}kov{\'a}} {\v{Z}}.,  {Chang} J.,   {Villarroel} B.,
  2020, \mn@doi [\aap] {10.1051/0004-6361/201936711}, \href
  {https://ui.adsabs.harvard.edu/abs/2020A&A...634A..66L} {634, A66}

\bibitem[\protect\citeauthoryear{{Martinez-Medina}, {Pichardo}, {Peimbert}  \&
  {Valenzuela}}{{Martinez-Medina} et~al.}{2019}]{Martinez-Medina2019}
{Martinez-Medina} L.,  {Pichardo} B.,  {Peimbert} A.,   {Valenzuela} O.,  2019,
  \mn@doi [\mnras] {10.1093/mnrasl/slz042}, \href
  {https://ui.adsabs.harvard.edu/abs/2019MNRAS.485L.104M} {485, L104}

\bibitem[\protect\citeauthoryear{{McGaugh}}{{McGaugh}}{2019}]{McGaugh2019}
{McGaugh} S.~S.,  2019, \mn@doi [\apj] {10.3847/1538-4357/ab479b}, \href
  {https://ui.adsabs.harvard.edu/abs/2019ApJ...885...87M} {885, 87}

\bibitem[\protect\citeauthoryear{{Monari}, {Famaey}  \& {Siebert}}{{Monari}
  et~al.}{2016a}]{Monari2016a}
{Monari} G.,  {Famaey} B.,   {Siebert} A.,  2016a, \mn@doi [\mnras]
  {10.1093/mnras/stw171}, \href
  {https://ui.adsabs.harvard.edu/abs/2016MNRAS.457.2569M} {457, 2569}

\bibitem[\protect\citeauthoryear{{Monari}, {Famaey}, {Siebert}, {Grand},
  {Kawata}  \& {Boily}}{{Monari} et~al.}{2016b}]{Monari2016b}
{Monari} G.,  {Famaey} B.,  {Siebert} A.,  {Grand} R. J.~J.,  {Kawata} D.,
  {Boily} C.,  2016b, \mn@doi [\mnras] {10.1093/mnras/stw1564}, \href
  {https://ui.adsabs.harvard.edu/abs/2016MNRAS.461.3835M} {461, 3835}

\bibitem[\protect\citeauthoryear{{Morgan}, {Sharpless}  \&
  {Osterbrock}}{{Morgan} et~al.}{1952}]{Morgan1952}
{Morgan} W.~W.,  {Sharpless} S.,   {Osterbrock} D.,  1952, \mn@doi [\aj]
  {10.1086/106673}, \href
  {https://ui.adsabs.harvard.edu/abs/1952AJ.....57....3M} {57, 3}

\bibitem[\protect\citeauthoryear{{Morgan}, {Whitford}  \& {Code}}{{Morgan}
  et~al.}{1953}]{Morgan1953}
{Morgan} W.~W.,  {Whitford} A.~E.,   {Code} A.~D.,  1953, \mn@doi [\apj]
  {10.1086/145754}, \href
  {https://ui.adsabs.harvard.edu/abs/1953ApJ...118..318M} {118, 318}

\bibitem[\protect\citeauthoryear{{Oort} \& {Muller}}{{Oort} \&
  {Muller}}{1952}]{Oort1952}
{Oort} J.~H.,  {Muller} C.~A.,  1952, Monthly Notes of the Astronomical Society
  of South Africa, \href
  {https://ui.adsabs.harvard.edu/abs/1952MNSSA..11...65O} {11, 65}

\bibitem[\protect\citeauthoryear{{Pantaleoni Gonz{\'a}lez}, {Ma{\'\i}z
  Apell{\'a}niz}, {Barb{\'a}}  \& {Reed}}{{Pantaleoni Gonz{\'a}lez}
  et~al.}{2021}]{Pantaleoni2021}
{Pantaleoni Gonz{\'a}lez} M.,  {Ma{\'\i}z Apell{\'a}niz} J.,  {Barb{\'a}}
  R.~H.,   {Reed} B.~C.,  2021, \mn@doi [\mnras] {10.1093/mnras/stab688}, \href
  {https://ui.adsabs.harvard.edu/abs/2021MNRAS.504.2968P} {504, 2968}

\bibitem[\protect\citeauthoryear{{Persic}, {Salucci}  \& {Stel}}{{Persic}
  et~al.}{1996}]{Persic1996}
{Persic} M.,  {Salucci} P.,   {Stel} F.,  1996, \mn@doi [\mnras]
  {10.1093/mnras/278.1.27}, \href
  {https://ui.adsabs.harvard.edu/abs/1996MNRAS.281...27P} {281, 27}

\bibitem[\protect\citeauthoryear{{Quillen}}{{Quillen}}{2002}]{Quillen2002}
{Quillen} A.~C.,  2002, \mn@doi [\aj] {10.1086/341379}, \href
  {https://ui.adsabs.harvard.edu/abs/2002AJ....124..924Q} {124, 924}

\bibitem[\protect\citeauthoryear{{Quillen}, {Minchev}, {Bland-Hawthorn}  \&
  {Haywood}}{{Quillen} et~al.}{2009}]{Quillen2009}
{Quillen} A.~C.,  {Minchev} I.,  {Bland-Hawthorn} J.,   {Haywood} M.,  2009,
  \mn@doi [\mnras] {10.1111/j.1365-2966.2009.15054.x}, \href
  {https://ui.adsabs.harvard.edu/abs/2009MNRAS.397.1599Q} {397, 1599}

\bibitem[\protect\citeauthoryear{{Quillen} et~al.,}{{Quillen}
  et~al.}{2018}]{Quillen2018}
{Quillen} A.~C.,  et~al., 2018, \mn@doi [\mnras] {10.1093/mnras/sty2077}, \href
  {https://ui.adsabs.harvard.edu/abs/2018MNRAS.480.3132Q} {480, 3132}

\bibitem[\protect\citeauthoryear{{Reid} et~al.,}{{Reid}
  et~al.}{2014}]{Reid2014}
{Reid} M.~J.,  et~al., 2014, \mn@doi [\apj] {10.1088/0004-637X/783/2/130},
  \href {https://ui.adsabs.harvard.edu/abs/2014ApJ...783..130R} {783, 130}

\bibitem[\protect\citeauthoryear{{Reid} et~al.,}{{Reid}
  et~al.}{2019}]{Reid2019}
{Reid} M.~J.,  et~al., 2019, \mn@doi [\apj] {10.3847/1538-4357/ab4a11}, \href
  {https://ui.adsabs.harvard.edu/abs/2019ApJ...885..131R} {885, 131}

\bibitem[\protect\citeauthoryear{{Roca-F{\`a}brega}, {Valenzuela}, {Figueras},
  {Romero-G{\'o}mez}, {Vel{\'a}zquez}, {Antoja}  \&
  {Pichardo}}{{Roca-F{\`a}brega} et~al.}{2013}]{Roca-Fabrega2013}
{Roca-F{\`a}brega} S.,  {Valenzuela} O.,  {Figueras} F.,  {Romero-G{\'o}mez}
  M.,  {Vel{\'a}zquez} H.,  {Antoja} T.,   {Pichardo} B.,  2013, \mn@doi
  [\mnras] {10.1093/mnras/stt643}, \href
  {https://ui.adsabs.harvard.edu/abs/2013MNRAS.432.2878R} {432, 2878}

\bibitem[\protect\citeauthoryear{{Roca-F{\`a}brega}, {Antoja}, {Figueras},
  {Valenzuela}, {Romero-G{\'o}mez}  \& {Pichardo}}{{Roca-F{\`a}brega}
  et~al.}{2014}]{Roca-Fabrega2014}
{Roca-F{\`a}brega} S.,  {Antoja} T.,  {Figueras} F.,  {Valenzuela} O.,
  {Romero-G{\'o}mez} M.,   {Pichardo} B.,  2014, \mn@doi [\mnras]
  {10.1093/mnras/stu437}, \href
  {https://ui.adsabs.harvard.edu/abs/2014MNRAS.440.1950R} {440, 1950}

\bibitem[\protect\citeauthoryear{{Sch{\"o}nrich}, {Binney}  \&
  {Dehnen}}{{Sch{\"o}nrich} et~al.}{2010}]{Schonrich2010}
{Sch{\"o}nrich} R.,  {Binney} J.,   {Dehnen} W.,  2010, \mn@doi [\mnras]
  {10.1111/j.1365-2966.2010.16253.x}, \href
  {https://ui.adsabs.harvard.edu/abs/2010MNRAS.403.1829S} {403, 1829}

\bibitem[\protect\citeauthoryear{{Shetty}, {Vogel}, {Ostriker}  \&
  {Teuben}}{{Shetty} et~al.}{2007}]{Shetty2007}
{Shetty} R.,  {Vogel} S.~N.,  {Ostriker} E.~C.,   {Teuben} P.~J.,  2007,
  \mn@doi [\apj] {10.1086/520037}, \href
  {https://ui.adsabs.harvard.edu/abs/2007ApJ...665.1138S} {665, 1138}

\bibitem[\protect\citeauthoryear{{Siebert} et~al.,}{{Siebert}
  et~al.}{2012}]{Siebert2012}
{Siebert} A.,  et~al., 2012, \mn@doi [\mnras]
  {10.1111/j.1365-2966.2012.21638.x}, \href
  {https://ui.adsabs.harvard.edu/abs/2012MNRAS.425.2335S} {425, 2335}

\bibitem[\protect\citeauthoryear{{Spekkens} \& {Sellwood}}{{Spekkens} \&
  {Sellwood}}{2007}]{Spekkens2007}
{Spekkens} K.,  {Sellwood} J.~A.,  2007, \mn@doi [\apj] {10.1086/518471}, \href
  {https://ui.adsabs.harvard.edu/abs/2007ApJ...664..204S} {664, 204}

\bibitem[\protect\citeauthoryear{{Vall{\'e}e}}{{Vall{\'e}e}}{2014}]{Vallee2014}
{Vall{\'e}e} J.~P.,  2014, \mn@doi [\aj] {10.1088/0004-6256/148/1/5}, \href
  {https://ui.adsabs.harvard.edu/abs/2014AJ....148....5V} {148, 5}

\bibitem[\protect\citeauthoryear{{Wegg}, {Gerhard}  \& {Portail}}{{Wegg}
  et~al.}{2015}]{Wegg15}
{Wegg} C.,  {Gerhard} O.,   {Portail} M.,  2015, \mn@doi [\mnras]
  {10.1093/mnras/stv745}, \href
  {https://ui.adsabs.harvard.edu/abs/2015MNRAS.450.4050W} {450, 4050}

\bibitem[\protect\citeauthoryear{{Weinberg}}{{Weinberg}}{1991}]{Weinberg1991}
{Weinberg} M.~D.,  1991, \mn@doi [\apj] {10.1086/170059}, \href
  {https://ui.adsabs.harvard.edu/abs/1991ApJ...373..391W} {373, 391}

\bibitem[\protect\citeauthoryear{{Willett} et~al.,}{{Willett}
  et~al.}{2013}]{Willett2013}
{Willett} K.~W.,  et~al., 2013, \mn@doi [\mnras] {10.1093/mnras/stt1458}, \href
  {https://ui.adsabs.harvard.edu/abs/2013MNRAS.435.2835W} {435, 2835}

\bibitem[\protect\citeauthoryear{{de la Vega}, {Quillen}, {Carlin},
  {Chakrabarti}  \& {D'Onghia}}{{de la Vega} et~al.}{2015}]{DelaVega2015}
{de la Vega} A.,  {Quillen} A.~C.,  {Carlin} J.~L.,  {Chakrabarti} S.,
  {D'Onghia} E.,  2015, \mn@doi [\mnras] {10.1093/mnras/stv2055}, \href
  {https://ui.adsabs.harvard.edu/abs/2015MNRAS.454..933D} {454, 933}

\bibitem[\protect\citeauthoryear{{van de Hulst}, {Muller}  \& {Oort}}{{van de
  Hulst} et~al.}{1954}]{vandeHulst1954}
{van de Hulst} H.~C.,  {Muller} C.~A.,   {Oort} J.~H.,  1954, \bain, \href
  {https://ui.adsabs.harvard.edu/abs/1954BAN....12..117V} {12, 117}

\makeatother
\end{thebibliography}

\bsp
\label{lastpage}

\end{document}